\documentclass[fleqn,usenatbib]{mnras}

\usepackage[T1]{fontenc}
\usepackage{ae,aecompl}

\usepackage{savesym}  %% KK added.
\usepackage{graphicx}	% Including figure files
\usepackage{amsmath}	% Advanced maths commands
\usepackage{amssymb}	% Extra maths symbols
\usepackage{natbib}
\savesymbol{iint}  %KK added
\usepackage{txfonts}  %KK added
\restoresymbol{TXF}{iint}  %KK added

\usepackage[british]{babel}             % British English hyphenation

\newcommand{\cii}{[C\,{\sc ii}] }

\title[A merger in the dusty, $z=7.5$ galaxy A1689-zD1?]{A merger in the
dusty, $z=7.5$ galaxy A1689-zD1?} 

\author[K.K. Knudsen et al.]{Kirsten K. Knudsen,$^{1}$\thanks{E-mail: kirsten.knudsen@chalmers.se}
Darach Watson,$^{2}$
David Frayer,$^{3}$
Lise Christensen,$^{2}$
\newauthor
Anna Gallazzi,$^{4}$ 
Micha{\l} J.~Micha{\l}owski,$^{5}$
Johan Richard,$^{6}$
Jes\'us Zavala$^{2,7}$
\\
$^{1}$Department of Earth and Space Sciences, Chalmers University
of Technology, Onsala Space Observatory, SE-43992 Onsala, Sweden \\
$^{2}$Dark Cosmology Centre, Niels Bohr Institute, University of Copenhagen,
Juliane Maries Vej 30, DK-2100 Copenhagen \O, Denmark \\
$^{3}$National Radio Astronomy Observatory, PO Box 2, Green Bank, WV 24944,
USA  \\
$^{4}$INAF-Osservatorio Astrofisico di Arcetri, Largo Enrico Fermi 5, I-50125
Firenze, Italy \\
$^{5}$SUPA (Scottish Universities Physics Alliance), Institute for Astronomy,
University of Edinburgh, Royal Observatory, Blackford Hill, Edinburgh EH9
3HJ, UK \\
$^{6}$Univ Lyon, Univ Lyon1, Ens de Lyon, CNRS, Centre de Recherche
Astrophysique de Lyon UMR5574, F-69230, Saint-Genis-Laval,  France
\\
$^{7}$Center for Astrophysics and Cosmology, Science Institute,
University of Iceland, Dunhagi 5, 107 Reykjavik, Iceland 
}

\date{Accepted 2016 November 23. Received 2016 November 21; in original form
2016 March 9}

\pubyear{2016}

\begin{document}
\label{firstpage}
\pagerange{\pageref{firstpage}--\pageref{lastpage}}
\maketitle

% Abstract of the paper
\begin{abstract}
%This is a simple template for authors to write new MNRAS papers.
%The abstract should briefly describe the aims, methods, and main results of the paper.
%It should be a single paragraph not more than 250 words (200 words for Letters).
%No references should appear in the abstract.
The gravitationally-lensed galaxy A1689-zD1 is one of the most distant
spectroscopically confirmed sources ($z=7.5$). It is the earliest known
galaxy where the interstellar medium (ISM) has been detected; dust emission
was detected with the Atacama Large Millimetre Array (ALMA). A1689-zD1 is
also unusual among high-redshift dust emitters as it is a sub-L* galaxy and
is therefore a good prospect for the detection of gaseous ISM in a more
typical galaxy at this redshift. We observed A1689-zD1 with ALMA in bands 6
and 7 and with the Green Bank Telescope (GBT) in band $Q$. To study the
structure of A1689-zD1, we map the mm
thermal dust emission and
find two spatial components with sizes about $0.4-1.7$\,kpc
(lensing-corrected). The rough spatial morphology is similar to what is
observed in the near-infrared with {\it HST} and points to a perturbed dynamical
state, perhaps indicative of a major merger or a disc in early formation. The
ALMA photometry is used to constrain the far-infrared spectral energy
distribution, yielding a dust temperature ($T_{\rm dust} \sim 35$--$45$\,K
for $\beta = 1.5-2$). We do not detect the CO(3-2) line in the GBT data with a
95\% upper limit of 0.3\,mJy observed. We find a slight excess emission in
ALMA band~6 at 220.9\,GHz. If this excess is real, it is likely due to
emission from the \cii\,158.8\,$\mu$m line at $z_{\rm [CII]} = 7.603$. 
The stringent upper limits on the \cii\/$L_{\rm FIR}$ luminosity ratio
suggest a \cii deficit similar to several bright quasars and massive
starbursts.  
\end{abstract}

% Select between one and six entries from the list of approved keywords.
% Don't make up new ones.

%%%%%%%%%%%%%%%%%%%%%%%%%%%%%%%%%%%%%%%%%%%%%%%%%%

%%%%%%%%%%%%%%%%% BODY OF PAPER %%%%%%%%%%%%%%%%%%
%________________________________________________________________
%
% % % % % % % % % % % % % % % % % % % % 
%   INTROUDCTION 
% % % % % % % % % % % % % % % % % % % % 

\section{Introduction}

With the increasing number of galaxies with spectroscopically confirmed redshifts $z>7$
\citep{vanzella11,ono12,schenker12,shibuya12,finkelstein13,oesch15,watson15,zitrin15,oeach16,song16}, 
the possibilities for quantifying and understanding the processes that take
place during the earlier stages of galaxy formation and evolution improve
significantly. 
Spectroscopic redshifts are necessary not only for determining the distance,
but also for enabling detailed follow-up observations.  
For the $z>6$ range, multiwavelength studies of quasars and submillimetre
galaxies have been very successful 
\citep[e.g.][]{maiolino05,venemans12,riechers13,wang13,willott15a,banados15,cicone15,venemans15},
however those sources represent the very bright end of the luminosity
function, and
they are not representative of the overall galaxy population. 

Recent searches for dust and far-infrared emission lines towards $z\sim7$
galaxies have resulted in only a few line detections
\citep[e.g.][]{kanekar13,ouchi13,gonzalez14,ota14,maiolino15,schaerer15,willott15b,knudsen16a,pentericci16,bradac16}. 
These galaxies have non-detections or marginal detections in the far-infrared
continuum suggesting that the dust-mass is relatively low. 
However, \citet{watson15}
found a clear detection of dust emission towards the spectroscopically
confirmed $z=7.5$ galaxy Abell\,1689-zD1 (A1689-zD1; \citealt{bradley08})
with an estimated dust mass comparable to that of the Milky Way.  
A1689-zD1 is strongly lensed by a factor of 9.3, thus providing
constraints on a sub-$L^\star$ galaxy (sub-$L^\star$ for $z\sim 7.5$).  
The estimated stellar mass is $\sim
1.7\times 10^{9}$\,M$_\odot$ and a total star formation rate (SFR) of
$\sim12$\,M$_\odot$\,yr$^{-1}$, which is dominated by dust-obscured star
formation \citep{watson15}.  Given the relatively short
time after the big bang, it is unclear what mechanisms have produced such a
large dust mass in this galaxy \citep[e.g.][]{michalowski15}.  

In this paper we present a follow-up study of A1689-zD1 aimed at a detailed
investigation of the structure of the dust emission and thus the distribution
of obscured star formation.   
Furthermore, we report observations of emission lines from \cii
1900.537\,GHz and CO(3-2) in order to study the properties of the
interstellar medium (ISM) and accurately measure the systemic redshift. 
In Section~\ref{sect:obs}, we present Atacama Large Millimetre Array (ALMA) and
Green Bank Telescope (GBT) observations.  In
Section~\ref{sect:res} we show the results of the observations and place it 
in the context of previous observations.  We discuss the implications of the
results in Section~\ref{sect:disc}.  
Throughout the paper we assume a $\Lambda$CDM cosmology with $H_0 =
67.3$\,km\,s$^{-1}$\,Mpc$^{-1}$, $\Omega_{\rm M} = 0.315$, and $\Omega_\Lambda =
0.685$ \citep{planck13}.  

%__________________________________________________________________
%
% % % % % % % % % % % % % % % % % % % % 
%   OBSERVATIONS 
% % % % % % % % % % % % % % % % % % % % 

\section{Observations}
\label{sect:obs}

\subsection{ALMA observations}

A1689-zD1 was observed with ALMA during Cycle-2 with the observations carried
out during December 2014 using band 6, and May 2015 and April 2016 with band
7. 
The purpose of the band-6 observations was to search for the redshifted \cii
line and, as the optical redshift is determined from the Lyman-$\alpha$ break
and not from emission lines \citep{watson15}, we
designed the observations to cover a large frequency range.  The receiver was
tuned with three different setups, each with three spectral windows in one
side band, which enabled continuous coverage from 216.51 to 231.06\,GHz
corresponding to the redshift range $z = 7.225-7.778$ for \cii\footnote{ 
This frequency range was selected based on the original error range to a
simple fit to the Lyman-$\alpha$ break in the X-Shooter spectrum.  However, a
more conservative estimate of the uncertainty based on multiple methods gives
a somewhat larger range and is quoted in \citet{watson15}.}.
The band 7 receiver was tuned to 343.5\,GHz for continuum measurements.  
The telescope configuration has baselines extending between 15 and 348\,m. 
Table \ref{tab:obssum} summarizes the details of the observations including
a list of the calibrators.

{\sc casa} ({\sc common astronomy software
application}\footnote{https://casa.nrao.edu}; \citealt{mcmullin07}) was used
for reduction, calibration, and imaging.  
The results from the pipeline reduction\footnote{For details on the pipeline,
see https://almascience.eso.org/documents-and-tools/} carried out by the
observatory was generally sufficient with only some minor extra flagging,
which did not change the final result.  
Exception to this was the flux calibration of the final band-7
observations, for which the flux calibration had to be adjusted with updated
flux density values for the calibrator J1256$-$0547 (a.k.a. 3C279).  This
update corresponded to about 10\%.  We note that according to the ALMA calibrator
source catalogue\footnote{\tt almascience.eso.org/sc/}, 3C279 was variable up
to 30\% in the month before our observations in 2016, though in April 2016
it appears to have been more stable.  While individual fluxes for 3C279 have
$\sim 5\%$ errors, we place a conservative flux calibration uncertainty of
10\% for our band-6 results. 
Continuum maps were made for both the band-6 and band-7 data.  
Using natural weighting, the obtained rms is 21\,$\mu$Jy\,beam$^{-1}$ and
45\,$\mu$Jy\,beam$^{-1}$, respectively.  
Additionally, a spectral cube was created for the band 6 data, and with natural
weighting in channels of 26\,km\,s$^{-1}$ width, the rms is $\sim
0.5-0.8$\,mJy\,beam$^{-1}$\,channel$^{-1}$ (typically the highest noise is
towards the highest frequency across the spectrum).  

Similar to our description in \citet{watson15}, the most conservative
estimate of the astrometric uncertainty is half the beam dimensions, meaning
$\sim 0.65$\,arcsec\,$\times 0.34$\,arcsec.  There is no notable emission detected from
the nearby, low-$z$ galaxy seen 1.5\,arcsec from A1689-zD1. 

We report the serendipitous detection of a background source in the band 6
data in a forthcoming paper (Knudsen et al., in preparation).

\begin{table*}
\caption[]{Summary of the ALMA observations, project 2013.1.01064.S.  Columns
are:  ALMA band number, date of observations, number of antennas
($N_{\rm ant}$), two columns of calibrators used for each set of
observations, and the central frequency of each spectral window except for
the continuum observations where we give the central frequency of the
continuum band. 
\label{tab:obssum} }
\begin{tabular}{lccccc}
\hline
\hline
& Date & $N_{\rm ant}$ & \multicolumn{2}{c}{Calibrators} & $\nu_{\rm spw, central}$ \\
 & (DD-MM-YYYY) & & Flux & Bandpass + Gain & (GHz) \\
\hline
{\it Band 6} & 08-12-2014 & 35 & Ganymede & J1256$-$0547 & 217.455, 219.036, 220.456 \\
& 08-12-2014 & 35 & Ganymede & J1256$-$0547 & 222.306, 223.806, 225.306 \\
& 08-12-2014 & 35 & Titan & J1256$-$0547 & 227.156, 228.606, 230.156 \\
{\it Band 7} & 05-05-2015 & 34 & J1256$-$0547 & J1256$-$0547 & 343.5 (continuum) \\ 
& 11-04-2016 & 42 & J1256$-$0547 & J1256$-$0547 & 343.5 (continuum) \\
\hline
\end{tabular}
\end{table*}

\subsection{GBT observations}

The GBT was used to search for the redshifted CO(3-2)
line ($\nu_{\rm rest} = 345.796$\,GHz).  The dual-beam $Q$-band receiver was
tuned to a central frequency of 40.265\,GHz covering the frequency range of
38.710-42.65\,GHz, which corresponds to $z_{\mathrm CO(3-2)} = 7.11 - 7.93$.  Five
observing sessions were carried out between 2014-09-18 and 2014-10-30.
Observing scans were taken with the 'SubBeamNod' mode which used the
subreflector to switch between the two beams every 6\,s.   We observed
the bright nearby quasar 3C279  every 50--60\,min to point and focus the
telescope and  to monitor the gain of the system.   
  
We used the new GBT spectrometer VEGAS for the spectral-line observations.
Four overlapping spectrometers (VEGAS mode-1) were used to cover the
bandwidth for each beam and for both circular polarizations with a raw
channel resolution of 1.465\,MHz. {\sc gbtidl} was used to carry out the data
reduction and  calibration.   The data have been corrected for the atmosphere
and losses due to drifts in pointing and focus during the observations.
Observations of 3C286 were used to derive the absolute calibration scale of
the data.   The uncertainty on the flux scale for the data is estimated to be
15\%.  For a channel resolution of 1.465\,MHz, we achieved $1\sigma$ rms of
0.267\,mJy, and smoothed to 8.79\,MHz (6 channels) the measured rms is
0.14\,mJy; 8.79\,MHz corresponds to 65\,km\,s$^{-1}$ for $\nu = 40.6$\,GHz.

%
% % % % % % % % % % % % % % % % % % % % 
%   RESULTS
% % % % % % % % % % % % % % % % % % % % 

\section{Results}
\label{sect:res}

\subsection{Continuum}
\label{subsect:cont}

\begin{figure}
\includegraphics[width=0.98\columnwidth]{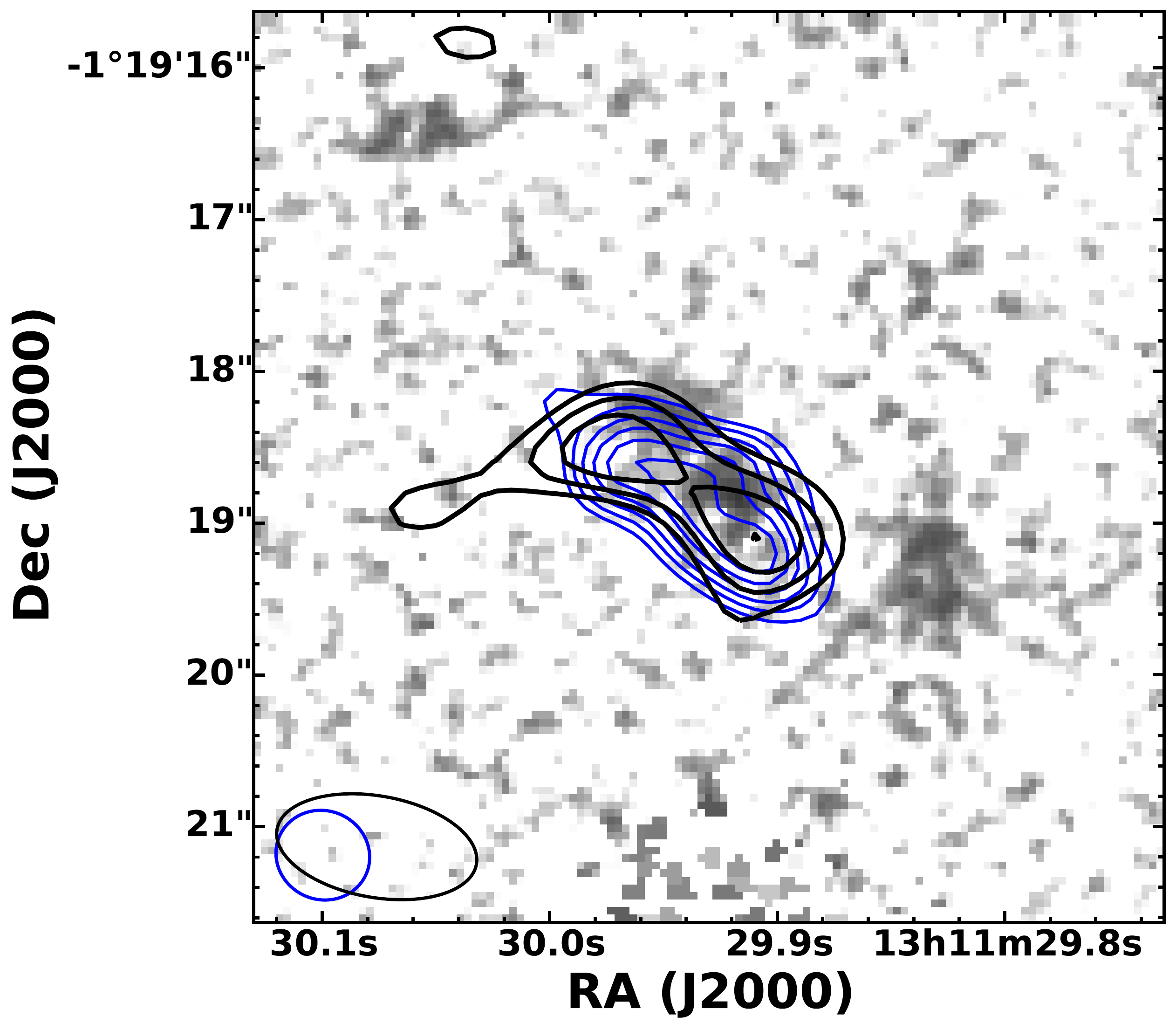}
\caption[]{
{\it HST} WFC3 F160W image overlaid with the contours of the ALMA band-6 and
-7
observations.  The ALMA band-6 continuum, shown in black, imaged with a
Briggs robust parameter of 0 resulting in an image resolution of
$1.33$\,arcsec\,$\times 0.67$\,arcsec (PA = 80 deg), which results in an rms of $\sigma =
30\,\mu$Jy.  The contours are in steps of 90, 120, 150, and 180\,$\mu$Jy.   
The blue contours show the band 7 observations, imaged using natural
weighting (resulting beam size of $0.62$\,arcsec\,$\times0.58$\,arcsec, PA =
67 deg) and with an rms of
$\sigma = 45\,\mu$Jy; contours are $3\sigma, 4\sigma, 5\sigma, 6\sigma,
7\sigma, 8\sigma$. 
\label{fig:images} }
\end{figure}

A1689-zD1 is detected in both bands 6 and 7 in continuum at $>10\sigma$.  
The emission is resolved with the structure showing two components. In
Fig.~\ref{fig:images} we show the bands 6 and 7 contours overlaid on an {\em
HST} near-infrared image.  Using the program {\sc uvmultifit} \citep{martividal14}
assuming a circular 2D Gaussian, we fit both bands together and measure a
diameter (FWHM) of $0.52\pm0.12$ arcsec and $0.62\pm0.12$\,arcsec for
the two components (NE and SW, respectively); in the fit we have allowed for
an astrometric offset between the two bands and find an insignificant offset
in right ascension of $0.008\pm0.108$\,arcsec, while in declination the
offset is $0.166\pm0.065$\,arcsec.  
Within the uncertainties, the two components have the same size. 
We note that fitting an elliptical 2D Gaussian function does not improve the
results and the estimated major--minor axis ratio is consistent with one
within the estimated uncertainties.
Lensing magnification\footnote{The estimates of the lensing magnification
are based on an updated mass model of Abell\,1689 \citep[][Richard et al.
in preparation]{limousin07} with the calculations done using {\sc lenstool}  
\citep{jullo07,jullo09}.}
is estimated to be $1.5\times6.5$ with the axis of the largest magnification
roughly along a position angle 90 degrees, which implies that the smallest scale
in that direction is about 0.4\,kpc.  This could indicate that the two clumps
are somewhat elongated (0.41--0.49\,kpc by 1.8--2.1\,kpc with an uncertainty
of about 50\%) as also seen in the lensing reconstruction of \citet{bradley08},
however, the ALMA data does not have sufficient quality to further reaffirm 
this.  The estimated area for both clumps together is $\sim 1$\,kpc$^{2}$,
consistent with the estimate based on the {\it HST} near-infrared data
\citep{watson15}.

From the {\sc uvmultifit} results we also obtain an estimate of the flux density
of $f_{\rm B7} = 1.33\pm0.14$\,mJy (the sum of $0.58\pm0.09$ and
$0.75\pm0.11$\,mJy, for NE and SW respectively)
and $f_{\rm B6} = 0.56\pm0.1$\,mJy (sum of $0.20\pm0.08$ and
$0.36\pm0.06$\,mJy for NE and SW respectively). The
errors reflect the uncertainties of the flux estimate and the errors on the
total flux density are obtained from adding the individual errors in
quadrature.

\begin{figure}
\includegraphics[width=0.98\columnwidth]{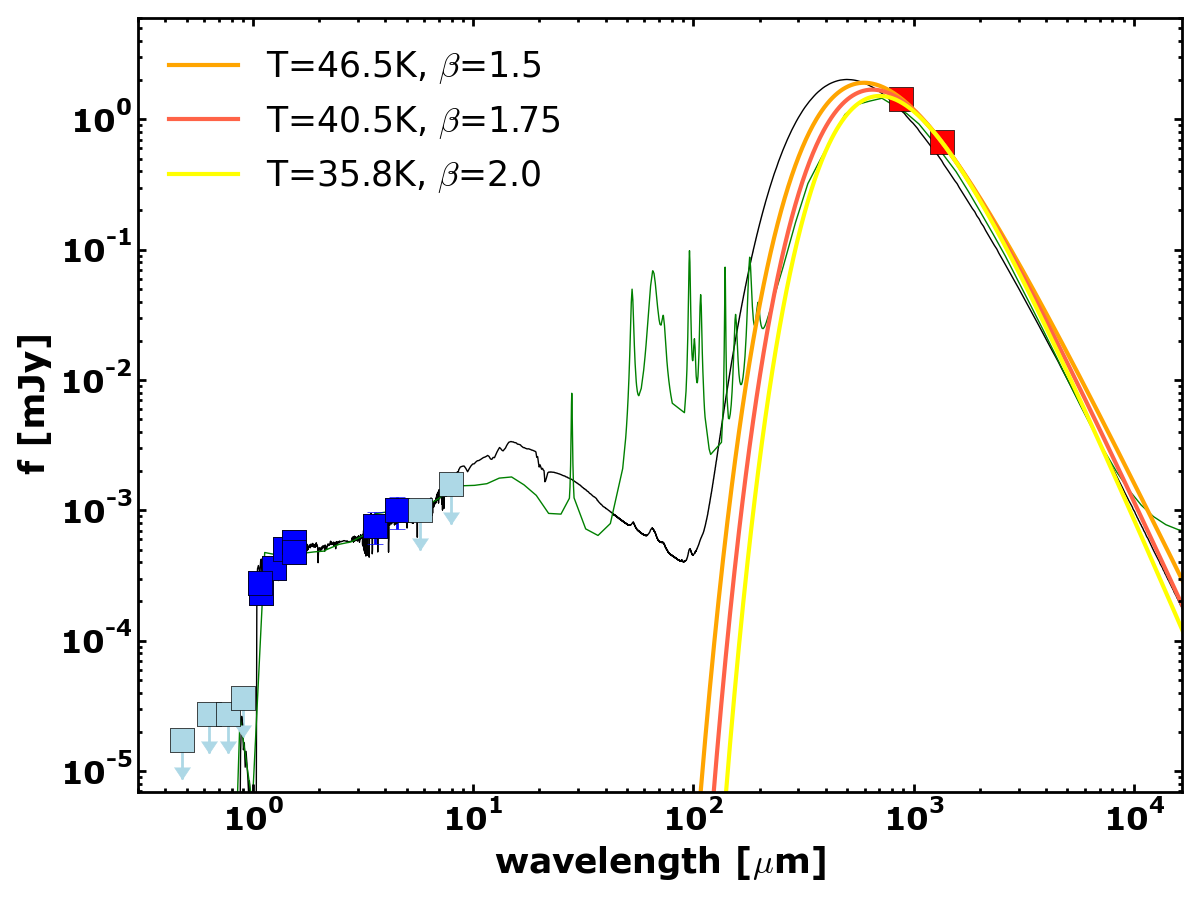}
\caption{The optical to mm SED of A1689-zD1.  The optical and
infrared measurements come from \citet{watson15} and \citet{bradley08}.   
The red, orange, and yellow lines show a modified blackbody for
the best-fitting temperature for varying $\beta$-values.  The ALMA photometry
has been corrected for the CMB effect according to the $\beta = 1.75$ and $T
= 40.5$\,K. 
The black and green lines shows the best-fitting SED models using {\sc magphys} and
{\sc grasil}. 
\label{fig:sed} }
\includegraphics[width=0.98\columnwidth]{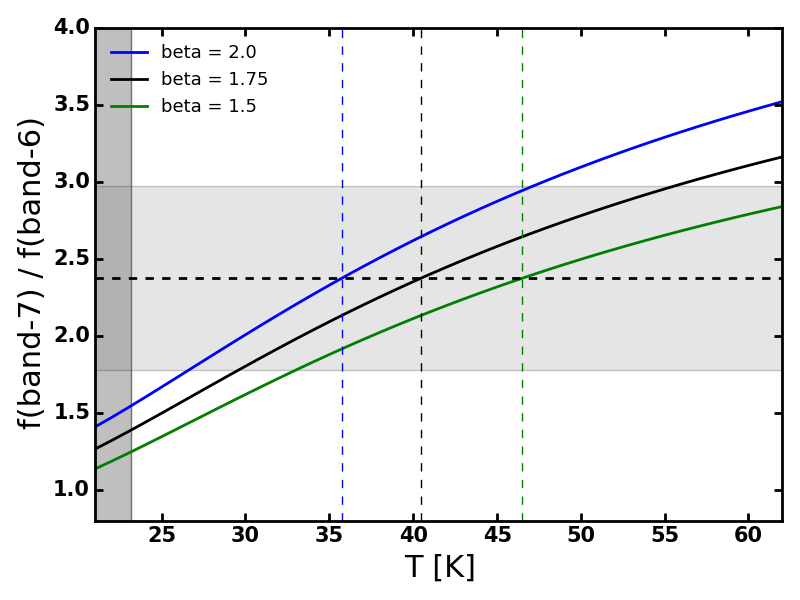}
\caption[]{The estimated flux ratio for the band 6 to band 7
ratio as a function of temperature for a modified blackbody spectrum
including corrections for the CMB radiation field; the errors on the two flux
estimates include both the uncertainty from the flux density estimate as well
as a 10\% uncertainty on the absolute flux calibration.  The horizontal dashed
line shows the measured ratio with the grey shaded area indicating the
error. 
The vertical grey shaded area towards low temperature indicates the $z=7.5$
CMB temperature.  
\label{fig:temp} }
\end{figure}

We use the photometric data points to constrain the far-infrared spectral
energy distribution (SED).  The
observed frequencies correspond to rest-frame wavelengths of 157.6\,$\mu$m and
102.7\,$\mu$m, respectively, for $z=7.5$.  This means we are probing close to
the peak of the dust emission.  
In Fig.~\ref{fig:sed} we show the measured fluxes together with the rest of
the SED as presented in \citet{watson15}.  
We constrain the temperature of a modified blackbody spectrum using the ratio
of the fluxes from bands 7 and 6, where the modified blackbody function
is described as $\nu^{\beta} B_\nu (T)$. 
For $z = 7.5$ the cosmic microwave background (CMB) temperature is $\sim
23.4$\,K.  We estimate that the CMB would increase the temperature by $\sim
1$\,per cent and that $\sim 5 - 23$\,per cent of the intrinsic flux is missed
due to the CMB radiation depending on the dust temperature, following
the analysis of \citet{dacunha13}.  When estimating the dust temperature of
A1689-zD1, we take 
the missed flux into account, but ignore the small change in temperature.  
In Fig.~\ref{fig:temp} we show how an estimated flux ratio changes with temperature
for three different $\beta$-values and compare this to our measured flux ratio
of $f_{\rm band-7} / f_{\rm band-6}  \sim2.4$.  We find the best-fitting values of
$T \sim 46.5$, 40.5, and 35.8\,K for $\beta = 1.5$, 1.75, and 2.0, respectively.
When not taking into account the effects of the CMB, the temperature
estimates would be about 5\,K higher.  
For $T=40.5$\,K and $\beta = 1.75$ the far-infrared luminosity is 
$1.7\times10^{12}$\,L$_\odot$ ($1.8\times10^{11}$\,L$_\odot$ after correcting
for magnification); we note that the fraction of missed flux is $\sim
8$ and $\sim 16$\,per cent for bands 7 and 6, respectively.   Using the
two other estimates for $\beta = 1.5$ and 2.0, the luminosity would
increase/decrease by about 20\,per cent.  
Given the uncertainties, this is in agreement with the estimate and
assumptions made in \citet{watson15}.  

We update the {\sc magphys} \citep{dacunha08} and {\rm grasil} 
\citep{silva98,iglesias07,michalowski10} SED model fits from \citet{watson15}
using the new ALMA band 6 + 7 photometry including the same correction for CMB effects as
described in the previous paragraph.  We find the derived parameters to be 
SFR = $12^{+4}_{-3}$\,M$_\odot$\,yr$^{-1}$, 
$\log ( M_{\rm stellar} / {\rm M}_\odot) = 9.3^{+0.13}_{-0.14}$ and 
$\log ( M_{\rm dust} / {\rm M}_\odot) = 7.6^{+0.27}_{-0.18}$ from {\sc
magphys}, and 
SFR = $14\pm8$\,M$_\odot$\,yr$^{-1}$, 
$\log ( M_{\rm stellar} / {\rm M}_\odot) = 9.4\pm0.1$, and
$\log ( M_{\rm dust} / {\rm M}_\odot) = 7.2\pm0.2$ from {\sc grasil}. 
The new and deeper ALMA measurements, provides a better constraint on the
far-infrared SED, e.g. for the dust temperature.  The largest change in the
derived parameters compared to \citet{watson15} is the lower nominal dust
mass from the {\sc grasil} fit, though this is consistent within the uncertainties.  
We plot the updated SED models in Fig.~\ref{fig:sed} and note that while the
models are quite similar, the largest difference is found on the Wien's tail
and mid-infrared range.  That is a wavelength range which is difficult to
cover with present instrumentation, however, part of it can be studied soon
with the James Webb Space Telescope. 

\begin{figure*}
\includegraphics[width=17.8cm]{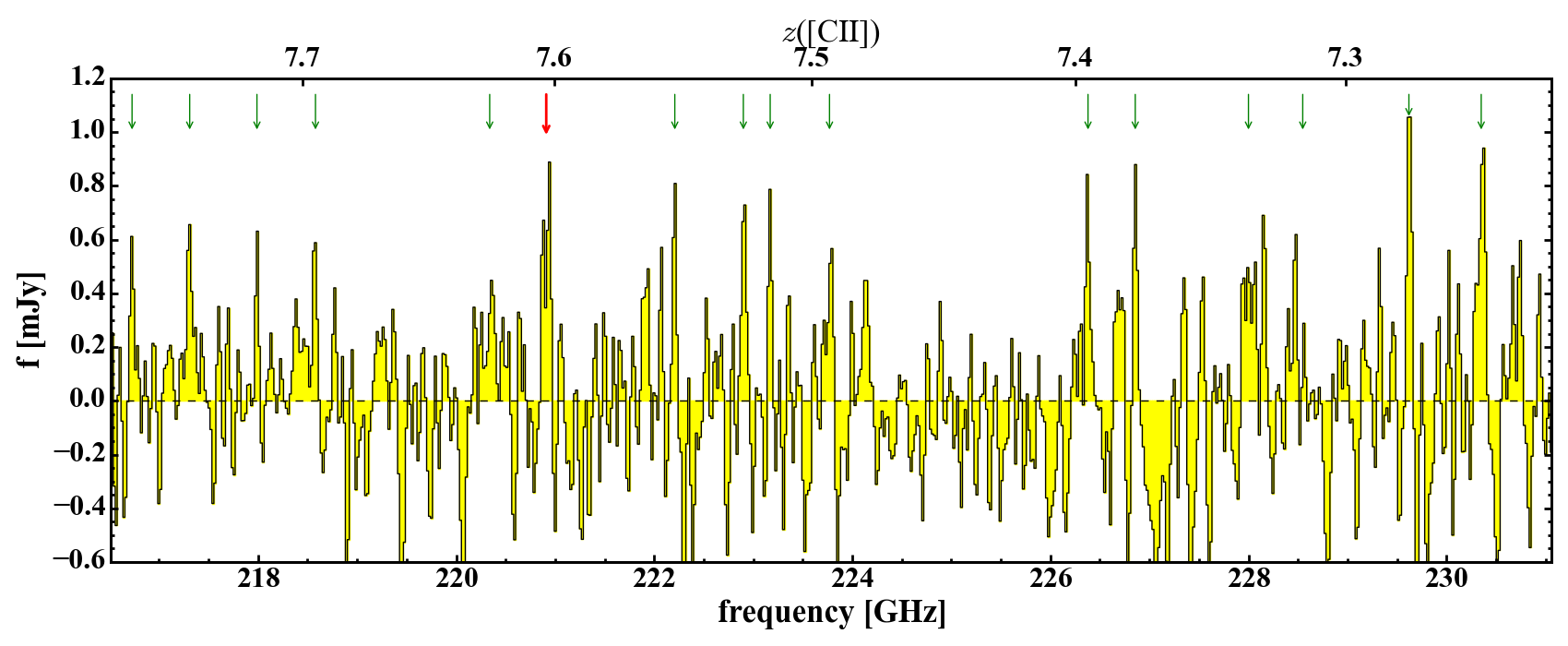}
\caption[]{The ALMA band 6 spectrum extracted at the position of the dust
continuum source.  The spectrum is binned to 26\,km\,s$^{-1}$ per channel,
Hanning smoothed by three channels, the continuum has been subtracted, and
the whole covered frequency range is shown.  The range corresponds to $z_{\rm
[CII]} = 7.2253 - 7.778$.  
The red arrow indicates the frequency of the tentative detection, see
Fig.~\ref{fig:tentline}, and the green arrows indicate the frequencies
investigated in the search for the \cii line; images are shown in Appendix
\ref{appsect:imaging}.  
\label{fig:result} }
\end{figure*}

\subsection{\cii line search} 
\label{subsect:cii}

\begin{figure}
\includegraphics[width=8cm]{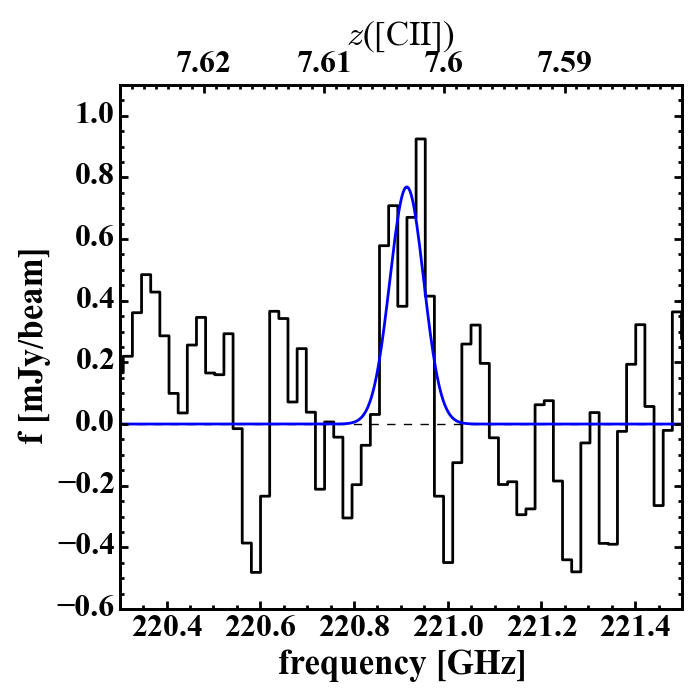}
\includegraphics[width=8cm]{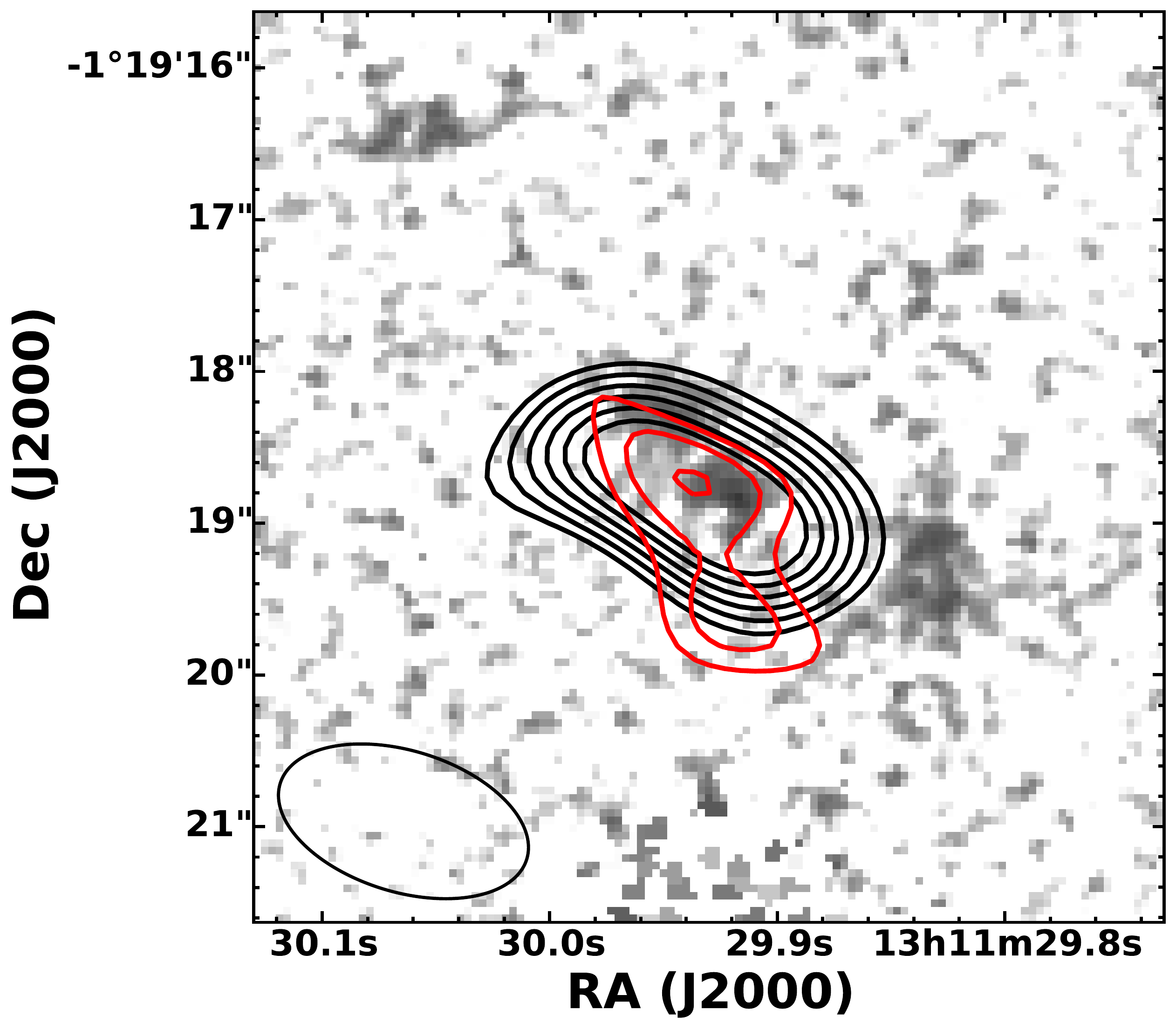}
\caption[]{Upper panel:  ALMA spectrum from Fig.~\ref{fig:result}, zoomed
to the {\it
tentative} detection at $\nu = 220.91$\,GHz.  The blue line shows the
Gaussian fit to the line, which yields a peak flux density of $S_{\rm peak} =
0.77\pm0.20$\,mJy\,beam$^{-1}$ and an integrated line intensity of $I_{\rm
[CII]} = 0.126\pm0.050$\,Jy\,km\,s$^{-1}$.   
Lower panel:  the integrated image is shown as red contours (2$\sigma$,
2.5$\sigma$, and
$3\sigma$) overlaid on the near-infrared image (see Fig.~\ref{fig:images})
with the black contours showing the continuum imaged with natural weighting
($5\sigma$, $6\sigma$, $7\sigma$, $8\sigma$, $9\sigma$, $10\sigma$);  the natural weighting produces an image with
lower resolution than e.g. is presented in Fig.~\ref{fig:images}.  
We note that the faint near-infrared source
eastwards of the ALMA contours is a low-redshift galaxy \citep{watson15}.
\label{fig:tentline} }
\end{figure}

We extract the spectrum from the band-6 data at the position of A1689-zD1 to
search for the \cii 1900.537\,GHz line, shown in Fig.~\ref{fig:result} (note
that the spectrum shown in the figure is Hanning smoothed and the r.m.s. is
$\sim0.3$\,mJy).  
The redshift range that the spectrum covers corresponds to $z_{\rm
[CII]} = 7.2253 - 7.778$.  
No strong emission line is detected.  
We investigate all tentative emission features above 0.6\,mJy (corresponding
to $2\sigma$; indicated in Fig.~\ref{fig:result}) and create maps integrated over 100\,MHz;  
the images are shown in Appendix~\ref{appsect:imaging}. 
At $\nu = 220.903$\,GHz we find a $3\sigma$ detection, which in spatial
distribution also would correspond to the position of A1689-zD1.  The
features at $\nu = 220.339$, 223.77, and 228.0\,GHz also show $3\sigma$
sources in the imaging, however, with a lower spatial overlap.  
If the spectral feature at $220.903$\,GHz is indeed a weak detection of the
\cii line, fitting a Gaussian
function implies a peak flux density of $S_{\rm peak} = 0.77\pm0.20$\,mJy\,beam$^{-1}$, 
FWHM line width of
$\Delta V = 163\pm49$\,km\,s$^{-1}$, and a velocity integrated intensity $I_{\rm [CII]} = 0.126 \pm
0.050$\,Jy\,km\,s$^{-1}$ (the error on the velocity integrated intensity is
derived from regular error propagation of the Gaussian fit), and a redshift
of $z_{\rm [CII]} = 7.6031\pm0.0004$.  We show the tentative line and the
image of the integrated line in Fig.~\ref{fig:tentline}. 
Under the assumption that this is indeed a detection of \cii, 
this would correspond to a line luminosity of $L_{\rm [CII]} \sim
1.7\times10^8$\,L$_\odot$ ($1.8\times10^7$\,L$_\odot$ after correcting for the
lensing magnification of $\mu = 9.3$) using 
$L_{\rm [CII]} = 1.04\times 10^{-3} S \Delta V D_{\rm L}^2 \nu_{\rm obs}$, 
where $S\Delta V$ is the velocity integrated flux density, $D_{\rm L}$ is the 
luminosity distance, and $\nu_{\rm obs}$ is the observer-frame frequency
\citep[e.g.][]{solomon05,carilliwalter13}.  
Given that this is only a possible detection, we use the line luminosity as
an upper limit.

\subsection{CO(3-2) line search}
\label{subsect:co32}

\begin{figure}
\includegraphics[width=0.99\columnwidth]{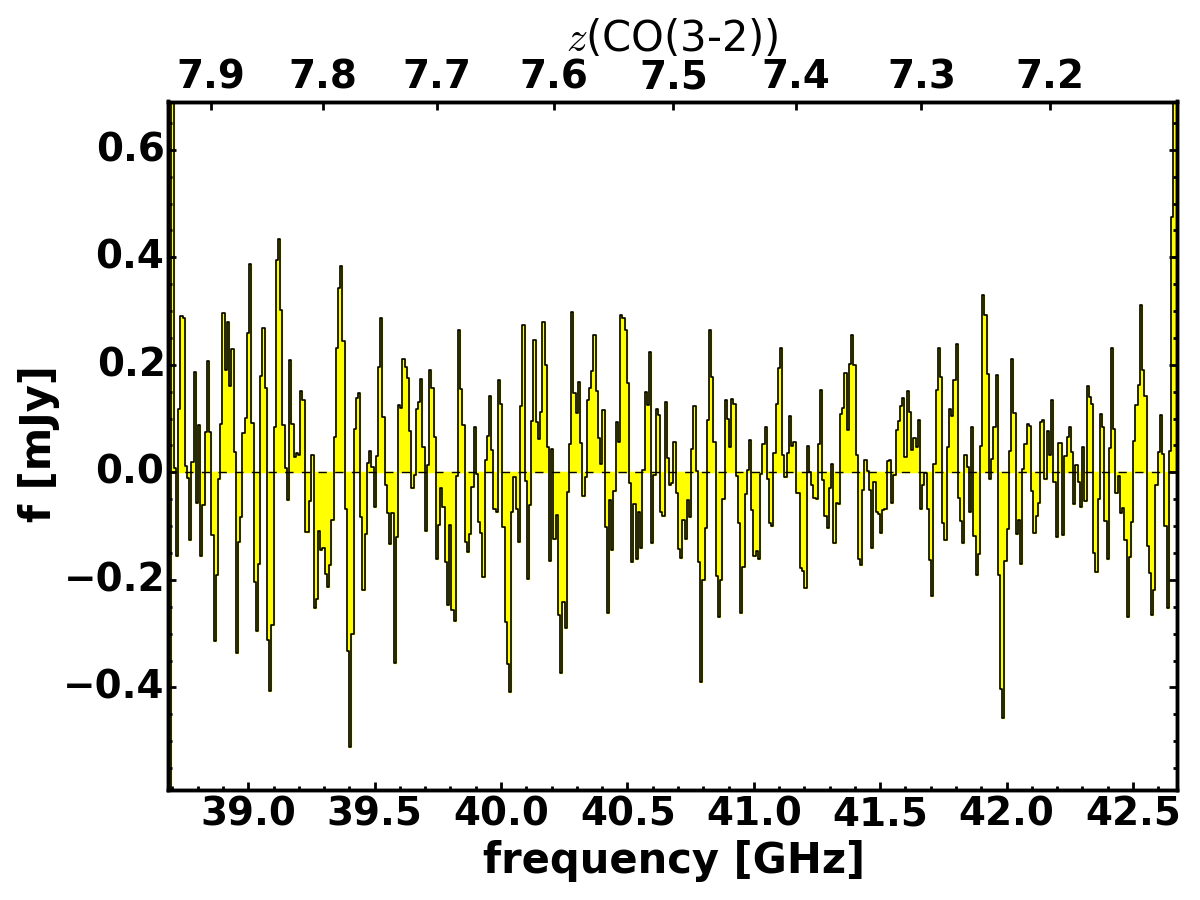}
\caption[]{The observed GBT spectrum searching for CO(3-2) covering a
redshift range of $z=7.11$--7.93.  For observed frequencies above
39.5\,GHz, the $1\sigma$ rms  is 0.14\,mJy for the 8.79 MHz ($\sim
65$\,km\,s$^{-1}$ at 40.5\,GHz) channels displayed here. The noise is
slightly higher at the edge of the receiver band below 39.5\,GHz. The
CO(3-2) emission line was not detected.
\label{fig:gbtspec} }
\end{figure}

In the GBT spectrum we do not detect the CO(3-2) line; the spectrum is shown in
Fig.~\ref{fig:gbtspec}. The redshift range
covered by the spectrum is $z_{\rm CO32} = 7.12 - 7.93$.  We place a
$3\sigma$ limit for the same linewidth as the tentative \cii line .  This 
corresponds to $L^{'}_{\rm CO32} = 8.6\times10^9$\,K\,km\,s$^{-1}$\,pc$^2$
($9.0\times10^8$\,K\,km\,s$^{-1}$\,pc$^2$ after correcting for lensing) and 
$L_{\rm CO32} = 1.1\times10^7$\,L$_\odot$ ($1.1\times10^6$\,L$_\odot$ after
lensing correction). 

Depending on the physical conditions of the gas, the CMB radiation 
field can significantly affect how well CO gas can be observed at very high
redshifts.  According to the estimates from \citet{dacunha13} for $z\sim
7.5$, the observed flux of CO(3-2) could be about half of the intrinsic flux
if the density and temperature are high ($T \sim 40$\,K, $n_{\rm H_2} \sim
10^{4.2}$\,cm$^{-3}$), and this ratio would be lower for lower temperature and
densities.  Thus the effect of the CMB could explain why we do not detect the
CO(3-2).  
We note that cosmic ray destruction of CO has been suggested as mechanism to
remove CO from the molecular gas mass \citep[e.g.][]{bisbas15}.  

Under the assumption that half of the intrinsic flux is missed due to the CMB
radiation field and that the CO(3-2) is thermally excited, we estimate an
upper limit on the 
molecular gas mass of $7.2\times10^9$\,M$_\odot$ using a conversion factor
$\alpha_{\rm CO} = 4$\,M$_\odot$\,K\,km\,s$^{-1}$\,pc$^2$
\citep[e.g.][]{carilliwalter13}.  Given the large number of assumptions, this
should be viewed as an order of magnitude estimate.  Using the size estimate
from the ALMA continuum observations, assuming a simple geometry, this would
imply an average gas density $<250$\,cm$^{-3}$.

%______________________________________________________________

% % % % % % % % % % % % % % % %
%   DISCUSSION 
% % % % % % % % % % % % % % % %

\section{Discussion}
\label{sect:disc}

A1689-zD1 is currently unusual among $z>7$ galaxies in that it is detected
in thermal dust emission and has a spectroscopically-confirmed redshift
without Ly$\alpha$ emission.  The observations presented here also show
that the galaxy is relatively small, with the galaxy resolved into two
separate components, both in thermal dust emission and stellar continuum.  
Despite its relative modest size, the mean stellar surface density
of about 1000\,M$_\odot$\,pc$^{-2}$ is not exceptionally high, however, the 
SFR surface density of $\sim
7$\,M$_\odot$\,yr$^{-1}$\,kpc$^{-2}$ as
traced from the dust emission, is similar to that of starbursts
\citep[e.g.][]{bouche07}; we note that the SFR surface density could be
partly overestimated if the size estimate from the submm imaging has been
underestimated because of the contrast against the CMB [this effect is
discussed in \citet{zhang16}]. 
Given the very high redshift of the galaxy, questions
naturally arise over how the galaxy came to be dusty and have apparently high
metallicity this early, its lack of strong UV and FIR emission lines, and
its morphology.

\subsection{Could A1689-zD1 be a merging galaxy pair?}

In order to assess the possibility of A1689-zD1 being a merging galaxy pair,
we estimate the descendant halo mass 
using the abundance matching technique, and use the results from $N-$body
cosmological simulations to approximate
the average merger rate. With a lensing-corrected UV luminosity of
$\sim1.8\times10^{10}$\,L$_\odot$, the mean halo mass
of A1689-zD1 is $\sim10^{11}$\,M$_\odot$ at $z=7.5$ (with a virial radius
$\sim11$~kpc). Specifically, we use the relation by \citet{schultz14}; 
see their Fig.~6.
This estimate is uncertain by a factor of $\sim2$, e.g. see Fig. 15 of
\citet{harikane15}. The
cumulative number of halo mergers with a mass ratio larger than $\xi_{\rm
min}$, between redshifts $z_0$ and $z$ that a
descendant halo of mass $M_0(z_0)$ has suffered is:
\begin{equation}\label{merger_rate}
	N_m(\xi_{\rm min}, M_0, z_0, z)=\int_{z_0}^z dz\int_{\xi_{\rm
min}}^1d\xi\frac{dN_m}{d\xi dz}[\left<M(z)\right>,\xi,z]
\end{equation}
where $dN_m/d\xi/dz$ is the mean merger rate in units of mergers per halo per
unit redshift per unit $\xi$, and $\left<M(z)\right>$ is the 
mean halo mass assembly history. In order to compute equation~\ref{merger_rate}, we
take the results of \citet{fakhouri10} 
based on the statistics of the Millennium II simulation (specifically, their
fitting formulae 1 and 2). Given the observed features of A1689-zD1,
we assume a major merger with a mass ratio of at least $\xi_{\rm min}=0.3$.
If we further take $z_0=7.5$ and $z=15$ (the maximum redshift where the
formulae can be trusted) in equation~\ref{merger_rate}, we find that $N_m\sim2.1$.
This number is most sensitive to the initial redshift $z$, e.g.
it drops by a factor of $\sim4$ if $z=9$ instead of $z=15$. 
Albeit given the evolved nature of the galaxy, it likely formed its first
stars earlier than $z\sim10$, thus $N_m$ is likely larger than 0.5. 
We can
then conclude that a merger under the conditions of A1689-zD1,  
should be a common occurrence according to our knowledge of structure
formation. With such a mass ratio, and a descendant mass, a merger 
will occur in a time-scale (governed by dynamical friction) of
$\mathcal{O}(100{\rm Myr})$ [\citet{binney87}; calibrated with the 
results of \citet{boylankolchin08}]. This time-scale is defined
from the time the virial radii of the two haloes started to overlap.

The clumpy structure of A1689-zD1 could possibly also be the result of
gravitational instabilities if A1689-zD1 is a disc galaxy in formation.  
Modelling by \citet{ceverino10} shows that giant clumps can arise in a
scenario where cold streams providing the gas can cause discs to be
gravitationally unstable and turbulent.  Signatures of clumps in disc
galaxies and 'proto-discs' are found both in optical studies of high-$z$
spiral galaxies \cite[e.g.][]{elmegreen09a,elmegreen09b} and in molecular gas
studies \citep[e.g.][]{tacconi10}, where the latter finds clumps with masses
$\sim 5\times10^{9}$\,M$_\odot$ and intrinsic radii $<1-2$\,kpc.  
We do not have the observational results to constrain whether A1689-zD1
is a merging galaxy pair or the result of cold mode accretion, however, the
important point is that the structure we observe is similar in both the
near-infrared (i.e.\ restframe UV) and thermal dust emission, indicative of
a dynamically pertubed system. 

In either case of interacting galaxies or clumpy structure arising from
dynamical instabilities, the
average density of the gas is increased and the SFR will increase.  
A1689-zD1 has a comparatively high SFR and dust continuum emission, and with a
stellar mass of $\sim 10^9$\,M$_\odot$ it is expected that the metallicity is
not low.  
These conditions seem to make A1689-zD1 suitable for accelerated grain 
growth, since undergoing a merger episode or dynamical instability could
enhance the gas density and thus potentially provide the necessary conditions
for grain growth in the ISM. 
For example, \citet{mancini15} present modelling where the time-scale for
ISM grain growth is inversely proportional to the density of the gas
and based on their model they suggest that the gas density of A1689-zD1 is
very high, possibly comparable to quasar host galaxies
\citep[e.g.][]{valiante14,mancini15}.  
However, given the uncertainties surrounding ISM grain growth
\citep[e.g.][]{2016MNRAS.463L.112F}, we note that conditions in the galaxy
certainly do not rule out direct SN dust production, and we await firmer
limits on dust destruction by SN shocks to decide this question one way or
the other.

\subsection{Redshift}

A1689-zD1 is one of the highest redshift galaxies known currently thanks to
the spectroscopic redshift determination.  The VLT X-Shooter spectrum
covered a wide wavelength range, both in the optical and near-infrared.  As
presented in \citet{watson15}, the redshift is determined from a clear break
around $\sim1\mu$m, though no lines were detected, resulting in an accuracy
of $\sigma_z \sim 0.2$.  One of the goals of the ALMA and GBT observations
was to observe line emission in order to get an accurate measurement of the
redshift.  As mentioned previously, both the ALMA and GBT data provide a wide
coverage for two lines that in lower redshift star-forming galaxies would be
bright.  We obtain a tentative $3\sigma$ detection of the \cii line with a
corresponding redshift of $z_{\rm [CII]} = 7.603$, however, deeper
observations are necessary to confirm this.  
It is possible that the systemic redshift of A1689-zD1 may not have been
covered by the observations. Adopting $\sigma_z \sim 0.2$ as a 68\% confidence interval,
which is very conservative, this corresponds to a minimum 80\% coverage of the
probability interval for the \cii line, and similarly a minimum 95%
coverage for the CO(3-2) line.  We will work in the next section on the
assumption that we cover these emission lines in these observations.
Equally, however, non-detections of these two lines could also
be due to astrophysical reasons in A1689-zD1 itself [see Sections~\ref{subsect:co32} and
\ref{subsect:gas}, also see e.g.\ \citet{maiolino15} for an example of $z>7$
galaxies with \cii non-detections]. We explore this possibility below.

\subsection{[C\,{\sc ii}]/$L_{\rm FIR}$ deficit?}
\label{subsect:gas} 

With the upper limit from the tentative detection of \cii we find that the
luminosity ratio is $L_{\rm [CII]}/L_{\rm FIR} < 0.0002$.  This is low
in comparison to local star-forming galaxies \citep[e.g.][]{malhotra01,diazsantos13}, which
typically have luminosity ratios of 0.0007--0.007 for galaxies with $L_{\rm
FIR} < 10^{11}$\,L$_\odot$.  In fact the limit is closer to the ratios
typically obtained for massive, luminous starbursts found in some
quasar host galaxies \citep[e.g.][]{wang13}, indicating that A1689-zD1 is
\cii deficient. 
While the reasons for this deficit observed towards
several massive starbursts remain unclear, a number of possible
explanations have been suggested \citep[e.g.][]{luhman03,stacey10}.  For
example, a high radiation field relative to the gas density could cause an
increased far-infrared luminosity compared to several emission lines such as
\cii \citep[e.g.][]{luhman03,abel09}.  If the gas density exceeds the
critical density of \cii\!, collisional de-excitation will become important and
reduce the cooling by \cii \citep[e.g.][]{goldsmith12}.  
Alternatively, if the gas temperature exceeds the excitation temperature of
the \cii line, a saturation of the upper fine-structure level would occur,
resulting in a maximum $L_{\rm [CII]}$ even at increasing $L_{\rm FIR}$ 
\citep[e.g.][]{munoz15}.   A recent model from \citet{narayanan16} suggests 
that the surface density plays an important role, i.e. an increase of the gas
surface density would lower the total amount of C$^+$ and an
increase of CO, which would result in a decreased $L_{\rm [CII]}/L_{\rm FIR}$
ratio.  

\begin{figure}
\centerline{\includegraphics[width=0.98\columnwidth]{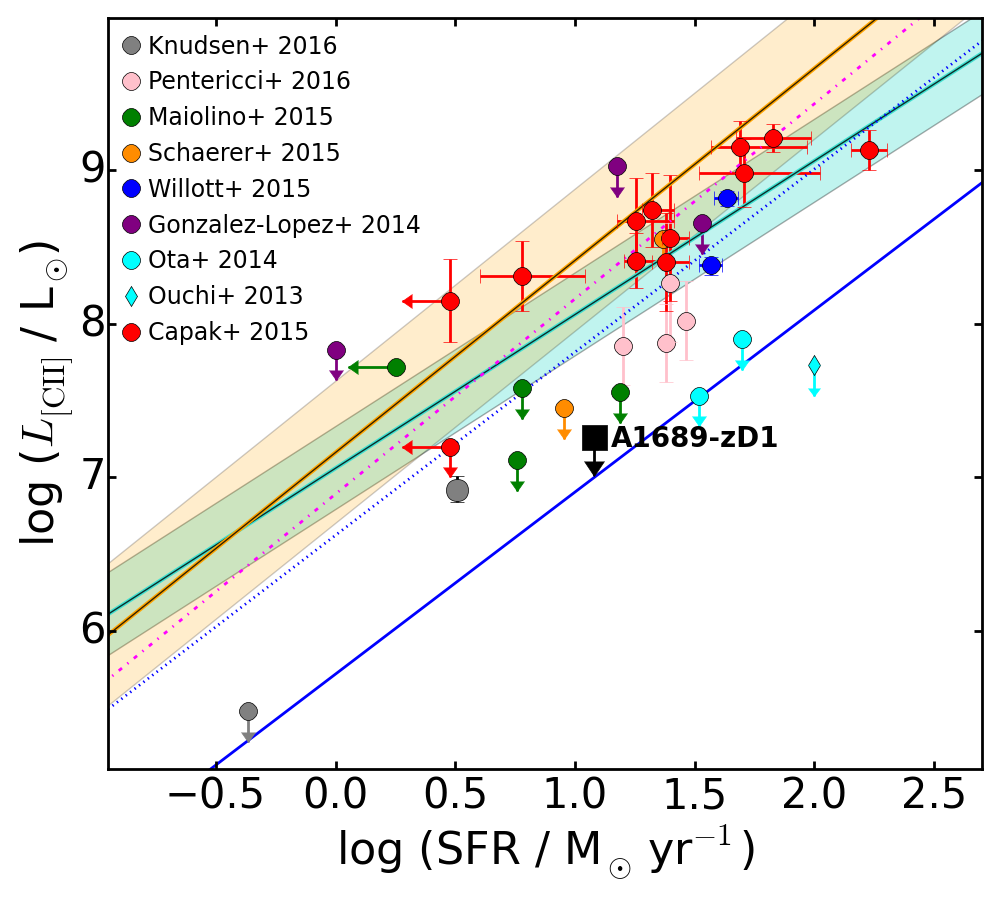}}
\caption[]{
We plot the \cii upper limit for A1689-zD1
after correcting for magnification (black square) against the star formation
rate. 
We also include the recent $z\sim 6$ results from
\citet{ouchi13}, \citet{gonzalez14}, \citet{ota14}, \citet{capak15},
\citet{maiolino15}, \citet{schaerer15}, \citet{willott14b},
\citet{knudsen16a}, and \citet{pentericci16}. 
The $L_{\rm [CII]}$--SFR relations from \citet{delooze14} are shown for 
local star-forming galaxies (turquoise line and region) and for 
low-metallicity dwarf galaxies (orange line and region).  
The resulting relation from the low-metallicity simulations from
\citet{vallini15} are indiciated with the blue lines (solid: $Z = 0.05
Z_\odot$, dotted: $Z=0.2Z_\odot$) and the magenta dash-dotted-line the results
for massive $z\sim2$ galaxies from
\citet{olsen15}.   
\label{fig:LciiSFR}}
\end{figure}

In nearby, star-forming galaxies, \cii has been found to be reliable tracer
of the SFR \citep[e.g.][]{delooze14,diazsantos14}.  
It has thus been expected that \cii
would be a bright tracer of the star formation taking place even in the
highest-$z$ galaxies, enabling discovery of sites of on-going (and obscured)
star formation as well as enabling means to reliably measure SFRs.  However,
the non-detections of non-SMG and non-QSO star-forming galaxies have revealed
a more complex picture of galaxy evolution during the first 0.5--1 Gyr after
the big bang.  In Fig.~\ref{fig:LciiSFR} we show the $L_{\rm [CII]}$--SFR
relation for $z > 6$ star-forming galaxies including A1689-zD1 in comparison
with the relations derived from low-$z$ galaxies. 

The lack of \cii detections
towards a large number of $z > 6$ star-forming galaxies has been discussed to
be the consequence of low metallicity
\citep{gonzalez14,ota14,capak15,maiolino15,schaerer15,willott15b,knudsen16a}. 
Low metallicity could bring down the
expected \cii line luminosity.   In fact modelling of the gas in $z>6$
galaxies suggests that the local SFR-$L_{\rm [CII]}$ relation is decreased
significantly depending on the metallicity \citep[e.g.][]{vallini15}.  
However, an explanation involving low metallicity in A1689-zD1 is hard to
reconcile with the high dust mass in this galaxy, a dust mass that suggests 
a metallicity close to the solar value \citep{watson15}. This makes the 
\cii\ deficiency here perplexing, and instead seems to point to a powerful 
radiation field or high gas densities as the likely culprits in depressing
\cii emission.  However, the measured SFR of $\sim12$\,M$_\odot$\,yr$^{-1}$
is below the characteristic mean SFR for this galaxy of
$\sim25$\,M$_\odot$\,yr$^{-1}$ (as calculated from the stellar mass divided
by the best-fitting age from the SED), and is not indicative of a galaxy at the
peak of a massive starburst event. This suggests that a powerful radiation
field is not the most obvious explanation for the \cii deficiency.
Our detection of a large dust-mass clearly demonstrates that this galaxy must
have a relatively high metallicity already.  

% % % % % % % % % % % % % % % %
%   CONCLUSIONS 
% % % % % % % % % % % % % % % %

\section{Conclusions}

In this paper, we present resolved observations of the dust continuum emission
from A1689-zD1, which is presently the most distant known galaxy ($z =
7.5\pm0.2$) with a detection of emission from thermal dust. 
Our findings are as follows: 
\begin{itemize}
\item The deep band 6 and band 7 continuum observations show that the redshifted
far-infrared emission is extended over two components, each of which has a FWHM
size of $\sim 0.45{\rm kpc}\times1.9$\,kpc (after correcting for the gravitational lensing). 
We note that the combination of ALMA sub-arcsec observations with gravitational
lensing provides an efficient approach to resolving the structure of the
dust-emitting region. 
The
gross far-infrared morphology is similar to the morphology of restframe UV light.
This suggests that the galaxy is either two proto-galaxies
interacting/merging or a clumpy proto-galactic disc.  
\item Based on bands 6 and 7 ALMA photometry we derive a dust temperature
$T\sim35-45$\,K ($\beta = 1.5-2$) after correcting for the CMB radiation field.
This implies a far-infrared luminosity of $1.8\times 10^{11}$\,L$_\odot$
(corrected for lensing magnification), in
agreement with our previous results based on single-band photometry.  
\item Based on deep ALMA band 6 spectroscopy with a 15\,GHz coverage, 
the \cii line is not detected.  We present a tentative $3\sigma$
detection, which would imply a systemic redshift of 7.603.  Using the derived
line intensity and luminosity of the tentative line, we find that the line is
underluminous relative to the far-infrared luminosity in comparison with
local normal star-forming galaxies.    Compared to the SFR, 
the upper limit is similar to the growing number of \cii non-detections in $z
> 6$ star-forming galaxies.  While the non-detection can be explained
astrophysically, we emphasize that the possibility for a different redshift
remains until a confirmed line-detection has been obtained.  
\item The CO(3-2) line is not detected in the GBT observations.  Given the
high temperature of the CMB radiation field at $z\sim 7.5$, it is difficult
to determine an upper limit, however, we place an estimate
on the limit of the molecular gas of $< 7\times10^{9}$\,M$_\odot$ (assuming a
Galactic $L'_{\rm CO}$-to-$M_{\rm H2}$ conversion).  
\end{itemize}
Previous models for grain growth show that the time-scale is inversely
proportional to the gas density.  
It is possible that the reported structure, indicative of galaxy interaction
or clumps, could be the signature of increased gas density and thus accelerated
grain growth.  
More importantly, the question remains whether the relatively large
dust mass is special for A1689-zD1 or if future, deeper observations of large
samples of galaxies will reveal a
larger population of dusty, normal galaxies at $z>7$.

\section*{Acknowledgements}
We thank the staff of the Nordic ALMA Regional Center node for their very
helpful support and valuable discussions on the ALMA data.  
We acknowledge the anonymous referee for useful suggestions for the manuscript. 
KK acknowledges support from the Swedish Research Council (grant No.:
621-2011-5372) and the Knut and Alice Wallenberg Foundation.  
LC is supported by YDUN grant No. DFF-4090-00079.
AG has been supported by the EU Marie Curie Integration Grant 'SteMaGE'  No.
PCIG12-GA-2012-326466.  
JR acknowledges support from the ERC starting grant CALENDS (336736). 
JZ is supported by the EU under a Marie Curie International Incoming
Fellowship, contract PIIF-GA-2013-62772.
  This paper makes use of the following ALMA data:
  ADS/JAO.ALMA\#2013.1.01064.S. ALMA is a partnership of ESO (representing
  its member states), NSF (USA) and NINS (Japan), together with NRC
  (Canada) and NSC and ASIAA (Taiwan) and KASI (Republic of Korea), in 
  cooperation with the Republic of Chile. The Joint ALMA Observatory is 
  operated by ESO, AUI/NRAO and NAOJ.

%-------------------------------------------------------------------

%\begin{thebibliography}{}
%\end{thebibliography}

\appendix

\section{Mapping of candidate emission lines}
\label{appsect:imaging}

\begin{figure*}
\includegraphics[width=17.8cm]{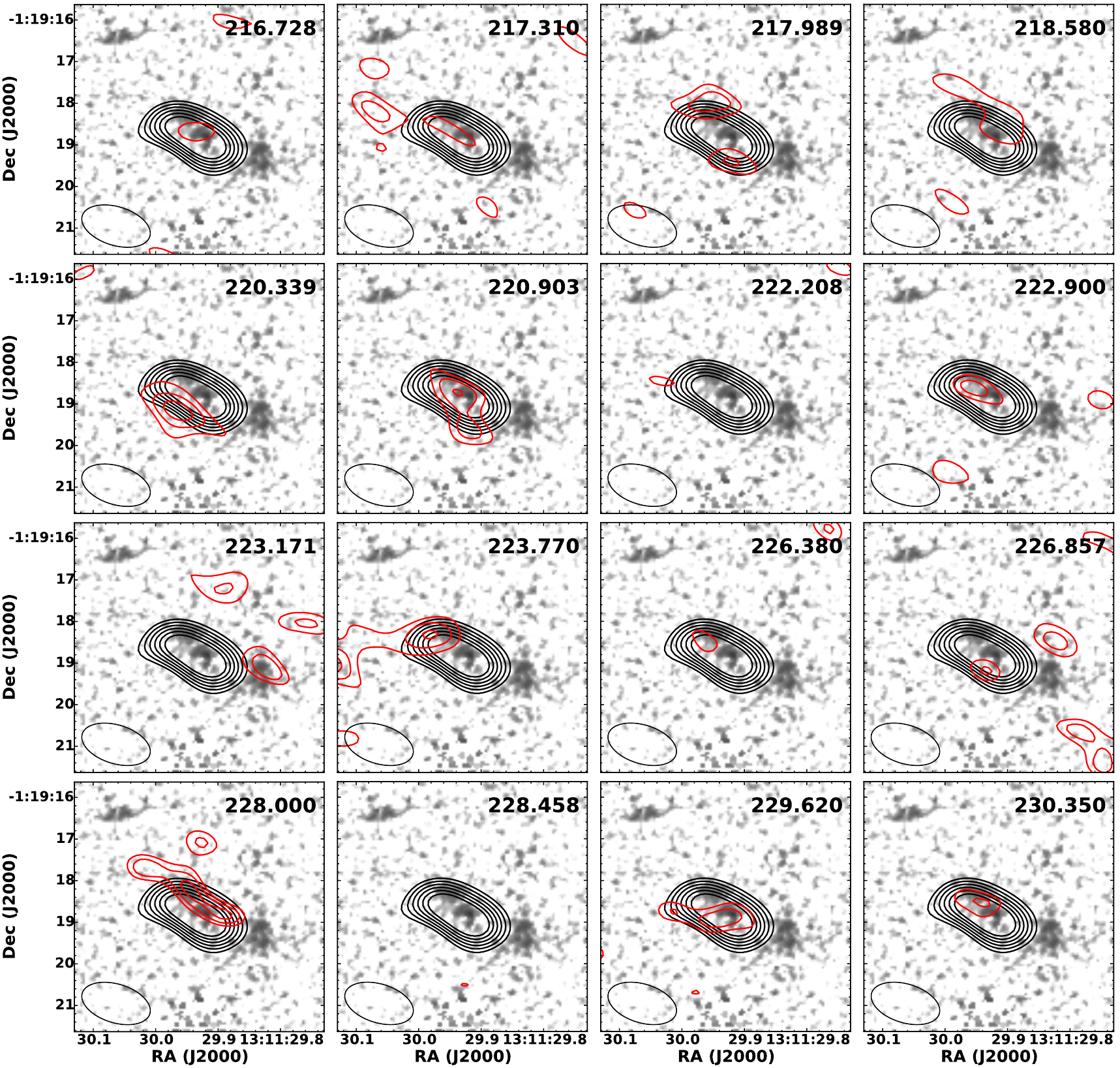}
\caption[]{Images from integrating over 0.1 GHz at different frequencies.
Red contours correspond to $2\sigma, 2.5\sigma$, and $3\sigma$, where
$\sigma$ is the r.m.s.  measured in each image.
The frequency is written in the upper-left corner of each sub-image and indicated
with a green arrow on top of the spectrum shown in Fig.~\ref{fig:result}. 
As in Fig.~\ref{fig:tentline}, the black contours show the continuum image with
natural weighting (5$\sigma$, 6$\sigma$, 7$\sigma$, 8$\sigma$, 9$\sigma$, and $10\sigma$); the grey-scale image is the
same as in Figs~\ref{fig:images} and \ref{fig:tentline}. 
\label{appfig:images}}
\end{figure*}

As described in Section~\ref{subsect:cii}, 
in order to search for the \cii line in the ALMA band 6 data, we imaged all
the positive spectral features with flux $>0.6$\,mJy, which corresponds to
$\sim 2\sigma$ in the Hanning-smoothed spectrum.   In this Appendix we show
the maps constructed using a width of 0.1\,GHz.  In Fig.~\ref{appfig:images}
we show the maps as contours overlaid on the near-infrared F160W {\it HST}
image.  
A couple of the features show image contours of $\sim
2.5\sigma$ around the position of A1689-zD1.  The features of 220.903,
223.770, and 228.000\,GHz show $3\sigma$ contours, however, only in the case of
220.903\,GHz are they following the same position as the dust continuum.  We
note that this is possibly the best candidate line for redshifted \cii and
therefore report it as a tentative detection. 

%%

%%%%%%%%%%%%%%%%%%%% REFERENCES %%%%%%%%%%%%%%%%%%

% The best way to enter references is to use BibTeX:

\bibliographystyle{mnras}
\bibliography{z7_alma}
%%%%%%%%%%%%%%%%%%%%%%%%%%%%%%%%%%%%%%%%%%%%%%%%%%

% Don't change these lines
\bsp	% typesetting comment
\label{lastpage}
\end{document}